\documentclass[11pt,a4paper]{article}

\pdfoutput=1
\usepackage{jheppub}

\usepackage{color}
\usepackage{appendix}
\usepackage{graphicx}
\usepackage{wrapfig,enumerate,slashed}
\usepackage{footmisc}
\usepackage{amsmath}
\usepackage{graphicx}
\usepackage{color}
\usepackage{comment}
\usepackage{hyperref}
\usepackage{epstopdf}
\usepackage{caption}
\usepackage{subcaption}

\DeclareMathAlphabet{\mathbold}{U}{zeur}{b}{n}

\renewcommand\({\left(}
\renewcommand\){\right)}
\renewcommand\[{\left[}
\renewcommand\]{\right]}

\newcommand{\refeq}{Eq.~}
\newcommand{\refsec}{Section~}
\newcommand{\reffig}{Fig.~}

\newcommand{\EH}{E_{\rm{H}}}
\newcommand{\lhx}{\lambda_{\rm{HX}}}
\newcommand{\lx}{\lambda_{\rm{X}}}
\newcommand{\lh}{\lambda_{\rm{H}}}
\newcommand{\mx}{m_{\rm{X}}}

\def\beq{\begin{equation}}
\def\eeq{\end{equation}}
\def\[{\begin{equation}}
\def\]{\end{equation}}

\begin{document}
\numberwithin{equation}{section}

\title{A Higgsploding Theory of Dark Matter}

\author[a]{Valentin V. Khoze,}
\author[a]{Joey Reiness,}
\author[a]{Jakub Scholtz}
\author[a]{and Michael Spannowsky}

\affiliation[a]{Institute for Particle Physics Phenomenology, Department of Physics, Durham University, DH1 3LE, UK}

\emailAdd{valya.khoze@durham.ac.uk}
\emailAdd{joey.y.reiness@durham.ac.uk}
\emailAdd{jakub.scholtz@durham.ac.uk}
\emailAdd{michael.spannowsky@durham.ac.uk}

\abstract{
We show that the Higgsplosion mechanism makes a prediction for the mass and coupling of a WIMP-like minimal scalar dark matter model. In particular the currently favoured minimal value for the Higgsplosion scale, $\EH \sim 25$ TeV, implies a dark matter mass $m_\mathrm{DM} \sim 1.25$ TeV and a moderate quartic coupling with the Standard Model Higgs field $\lambda_\mathrm{H,DM} \sim 0.4$. This point in the parameter space is still allowed by all current experimental bounds, including direct detection (XENON), indirect detection (HESS, Fermi, Planck) and collider searches. We have updated the scalar dark matter bounds to reflect the latest results from XENON and HESS experiments. We also comment on vacuum stability and dark matter self-interactions in this model. 
}

\date{}

\preprint{IPPP/18/16}

\maketitle

\section{Introduction}
\label{sec:intro}
\medskip

It is rather apparent that there is a dark matter (DM) component in our Universe. The earliest evidence comes from the rotation curves of galaxies pioneered by Vera Rubin \cite{1970ApJ...159..379R}. There is additional evidence in the power spectrum of the cosmic microwave background (CMB) and large scale structure, as seen most recently by the Planck satellite \cite{2016A&A...594A..13P} and the BOSS collaboration \cite{2017MNRAS.470.2617A}. There are geometrical proofs for existence of dark matter \cite{2002ApJ...577..183B} inside individual galaxies that do not rely on kinematic information alone. Finally, there are also colliding clusters of galaxies, such as the Bullet Cluster, that show the need for a non-baryonic matter component in our Universe \cite{Clowe:2003tk,Markevitch:2003at}.   

Unfortunately, all evidence for dark matter thus far has been of gravitational origin: non-gravitational evidence continues to elude us. Without any such additional evidence, the range of models that  both successfully provide the source of dark matter and evade all present constraints is large. The standard DM candidate is a weakly interacting massive particle (WIMP), but there are a plethora of other solutions, such as: axions, fuzzy dark matter, light sterile neutrinos, self-interacting dark matter, dissipative dark matter, atomic dark matter and many more.

In this article we focus on a classic WIMP-like example: a massive scalar whose $\mathcal{O}(1)$ coupling to the Higgs boson naturally generates a correct relic abundance $\Omega_{DM}h^2 = 0.12$ \cite{2016A&A...594A..13P}. This scenario has been well explored by many authors \cite{PhysRevD.50.3637,BURGESS2001709,PhysRevD.79.015014,Cline:2013gha,Escudero:2016gzx}.

Typically, the addition of a light fundamental scalar in a theory introduces hierarchy problems, and so previous scalar dark matter models suffered from inability to predict the preferred mass range for DM. However, in the Higgsplosion scenario \cite{Khoze:2017tjt, Khoze:2017lft, Khoze:2017uga}, this large hierarchy problem is greatly reduced by the presence of a new dynamically-generated scale -- the Higgsplosion scale. This scale determines corrections to the DM mass. As a result, if the scale is known, we obtain a unique prediction for the mass of minimal scalar dark matter.

Currently the Higgsplosion scale is quantitatively not well known, although existing calculations indicate $\EH \sim 200 m_h = 25$ TeV 
\cite{Khoze:2017ifq} (see also Refs.~\cite{Libanov:1994ug,Gorsky:1993ix,Son:1995wz,Jaeckel:2014lya} for earlier work).
Experimentally, a low Higgsplosion scale can result in striking signatures at high-energy colliders \cite{Gainer:2017jkp}. In this paper we treat the Higgsplosion scale as a free parameter and show that the currently theoretically preferred region for the Higgsplosion scale leads to a dark matter model that is: a) not in tension with any current experimental results, and b) testable in the foreseeable future at direct detection experiments such as LZ and indirect detection experiments such as CTA.

This paper is organized as follows: in \refsec\ref{sec:pheno} we describe the model, clarify the scalar mass-Higgsplosion scale relation and enforce the freeze-out constraints. In \refsec\ref{DMD} we investigate the viability of such a theory, given both current and projected future constraints from direct detection, indirect detection and production experiments, including XENON and HESS. The strength and relevance of self-interaction and the stability of the vacuum for such a model are discussed in \refsec\ref{sec:other}. Finally, we present our concluding remarks in \refsec\ref{sec:concl}.

\medskip
\section{A minimal scalar dark matter model}
\label{sec:pheno}

\subsection{Model}
\label{sec:model}

In this work, we consider the simplest possible scenario for Higgsploding dark matter: a singlet real scalar, $X$, with a $\mathbb{Z}_2$ symmetry \cite{BURGESS2001709,PhysRevD.50.3637,PhysRevD.79.015014}. The dark and Standard Model sectors communicate via a Higgs-portal coupling, $\lhx$,
\begin{equation}
\mathcal{L} = \mathcal{L}_{SM}+ \frac{1}{2}\partial_\mu X \partial^\mu X - \frac{1}{2}m_{X,0}^2 X^2 - \frac{\lx}{4!} X^4 - \frac{\lhx}{2} X^2 \left(H^\dagger H\right) ,
\end{equation}
with,
\begin{equation}
\mathcal{L}_{SM}\supset \mu_0^2H^\dagger H - \lh(H^\dagger H)^2.
\end{equation}

For now, we make the assumption that $\lx\ll\lhx$. In \refsec \ref{sec:vac}, we confirm this is a safe assumption, despite the renormalization group (RG) flow. We also assume that the bare mass is small, such that the renormalized mass of the dark scalar is dominated by the quadratically-divergent contribution from loops of Higgs particle,
\begin{equation}
	\mx^2 = m_{X,0}^2+\delta \mx^2\approx\delta \mx^2\approx \frac{\lhx \Lambda_{\rm{UV}}^2}{16\pi^2},
\end{equation}
where $\Lambda_{\rm{UV}}$ is the UV cut-off of the theory. In a Higgsploding theory, as discussed in \cite{Khoze:2017ifq,Khoze:2017lft,Khoze:2017tjt,Khoze:2017uga}, this cut-off becomes physical. Above a certain virtuality, called the Higgsplosion scale $\EH$, the Higgs bosons are expected to decay exponentially into a large number of soft quanta, a phenomenon dubbed Higgsplosion. The imaginary part of the self-energy for the Higgs particle grows exponentially with the virtuality of the propagator and the Higgs propagator effectively vanishes above $\EH$. This has the consequence of cutting off integrals over Higgs four-momenta, $k^\mu$, at $k^\mu k_\mu=\EH^2$. Hence, in the regime described, we expect a dark scalar mass of order
\begin{equation}
\mx^2 \approx \lhx\frac{\EH^2}{16\pi^2}.
\label{eq:massscale}
\end{equation}
If the Higgsplosion scale is known, this relation greatly restricts the parameter space available, and uniquely determines the dark matter mass when combined with the freeze-out condition, as we will show in \refsec\ref{sec:freeze}.

As we will see, the region of parameter space corresponding to large mass and portal coupling remains unbounded for our model. As a result, we must choose a maximal coupling strength that we  consider perturbatively under control. In this work, we choose a somewhat conservative value, $\lx,\lhx \leq \sqrt{8\pi} \simeq 5$\cite{Panico:2015jxa}.

\subsection{Bare masses, scales and hierarchy}
\label{sec:masses}

Typically, in order to arrive at the IR spectrum we observe for the SM, we must fine-tune the Higgs bare mass in the UV -- this is the Hierarchy problem. This problem persists in the presence of Higgplosion, but is significantly reduced. In order to achieve the observed Higgs mass, the bare mass must now instead satisfy,
\begin{equation}
\mu_0^2 = -\frac{\lh}{2} v_{EW}^2 - \EH^2 \( \frac{\lh}{4 \pi^2}  +\frac{\lhx}{16 \pi^2} - \frac{N_c y_t^2}{8\pi^2} + \cdots \), 
\end{equation}
where the first term is the Higgs doublet mass required to break the electroweak symmetry and the remaining terms are self-energy corrections, with $N_c$ quark colours.
In the Higgsploding regime, we expect the scale of all of these contributions to be proportional to $\EH^2$: if any particle reaches virtuality of order $\EH^2$, it's self-energy quickly suppresses its propagator.  As a result, the Higgs mass\footnote{The top quark contribution is also reduced by Higgsplosion, we refer the reader to references~\cite{Khoze:2017lft, Khoze:2017ifq, Khoze:2017tjt,Khoze:2017uga} } is fine-tuned to the extent of $\sqrt{\lh}v_{\rm{EW}}/\EH \sim 10^{-2}$, which is a vast improvement when compared to the usual fine-tuning of order $m_h/m_{\rm GUT} \sim 10^{-14}$.

Given that $X$ is also a scalar, one might expect it to exhibit its own Higgsplosion-like behaviour, which we dub ``Xplosion''\footnote{This ``pun'' being the sole reason for our particle naming scheme.}. Indeed, in the limit $\lhx\to 0$, the dark sector decouples from the SM sector and one could expect the amplitude for the process 
\[
 X\to n X,
\label{eq:Xpl}
\] 
to grow exponentially at some Xplosion scale, $E_\mathrm{X} \propto f(8\pi/\lx) \mx$, where 
$f(8\pi/\lx)$ becomes infinite in the limit $\lx \to 0$.
In the symmetric phase of the theory, the processes  \eqref{eq:Xpl} most likely remains negligible.
It is only in the broken phase, i.e. for the scalar $X$ model with the spontaneously broken $\mathbb{Z}_2$ symmetry, that the calculations of
quantum effects summarised  in \cite{Khoze:2017ifq, Khoze:2017uga} would lead to Xplosion. However, in the $\mathbb{Z}_2$ symmetric case relevant here, where $X$ is the scalar DM candidate, quantum effects are known to exponentially suppress the effect of Xplosion seen at the classical level.
Furthermore, we have assumed that $\lhx\neq 0$ and $\mx\gg m_h$, so the {\it Higgs}plosion process,
\[
X\to X+nh,
\label{eq:Hpl}
\]
is allowed at far lower virtualities of order $\mx^2+\EH^2 \ll E_X^2$. Even if Xplosion \eqref{eq:Xpl} was possible, 
Higgsplosion \eqref{eq:Hpl} ultimately determines the UV behaviour of $X$.

In our scenario we set the bare mass of X to be relatively small,
\begin{equation}
- \frac{\lhx}{16\pi^2}\EH^2 \lesssim m_{X,0}^2 \lesssim \frac{\lhx}{16\pi^2}\EH^2,
\end{equation}
so that the self-energy contribution is dominant. In that sense, the mass of $X$ is natural.

\subsection{Freeze-Out}
\label{sec:freeze}
In this subsection, we follow treatment of freeze out calculations as summarised in \cite{Kolb:1990vq,BURGESS2001709}. An important quantity in the calculations to follow is the thermally-averaged dark matter annihilation cross-section $\langle \sigma_{ann}v_{rel}\rangle \equiv \langle \sigma v\rangle$. The contributions to this cross-section for different annihilation modes were calculated in \cite{PhysRevD.50.3637}. We include the exact results in Appendix~\ref{app:xsec}.

In the regime $\mx \gg v$, where $v$ is the Standard Model Higgs vacuum expectation value, the annihilation cross-section is dominated by the $hh$, $W_L^+W_L^-$ and $Z_L^0Z_L^0$ modes. This is not surprising: these modes are the four degrees of freedom of the Higgs doublet that directly couple to $X$ through the $\lhx X^2 H^\dagger H$ operator.
In the regime $\mx^2 \gg \{v^2,m_h^2,m_W^2,m_Z^2\}$ the annihilation cross-section simplifies to:
\begin{equation}
\langle\sigma v\rangle\approx \frac{\lambda^2_{HX}}{16\pi \mx^2} .
\label{eq:approx}
\end{equation}
We will see that $\mx \lesssim v$ leads to values for the Higgsplosion scale that are too low, and so we can safely ignore the low-mass regime entirely. As a result,  we will use the approximation of \refeq\ref{eq:approx} throughout this paper.

The relic density of dark matter today is constrained by the Planck satellite  \cite{2016A&A...594A..13P} to $\Omega_{\rm DM}h^2 = 0.12$.  Our prediction for the present day density of $X$ particles is \cite{BURGESS2001709, Kolb:1990vq,Dodelson:1282338}:
\begin{equation}
\Omega_\mathrm{X}h^2 = \left[  \frac{8\pi G g_*(\mx/x_f)}{45} \right]^{1/2} \frac{ 4\pi^2 G x_f T_0^3}{45\langle \sigma v \rangle  H_{100}^2},
\label{eq:freezeout}
\end{equation} 
where $g_*(\mx/fx_f)$ denotes the number of degrees of freedom in equilibrium at annihilation, $T_0$ is the current temperature, $H_{100}=100$km s$^{-1}$Mpc$^{-1}$, and $G$ is the Newton's constant. We determine the inverse freeze-out temperature, $x_f=\mx/T_F$, by solving the implicit equation,
\begin{equation}
x_f \sim \ln \left[ \sqrt{\frac{45}{4\pi^5}} \frac{1}{\sqrt{g_*}} \frac{m}{\sqrt{8\pi G}} \langle\sigma v\rangle \right] - \frac{1}{2}\ln^2\left[  \sqrt{\frac{45}{4\pi^5}} \frac{1}{\sqrt{g_*}} \frac{m}{\sqrt{8\pi G}} \langle\sigma v\rangle \right].
\label{eq:imp}
\end{equation}
Solving the above equation for a perturbative range of $\lhx$ yields the usual $x_f$ values in the range $20$--$30$. The freeze-out condition in \refeq(\ref{eq:freezeout}) gives a relationship between $\lhx$ and $\mx$:
\begin{equation}
\lhx = 0.30 \(\frac{x_f}{20}\)^{1/2}\(\frac{\Omega_\mathrm{X}h^2}{0.12}\)^{1/2}\(\frac{\mx}{1\mathrm{ TeV}}\).
\end{equation}

The combination of imposing the Higgsploding scenario, keeping $\lhx$ perturbative and demanding the correct relic density greatly limits the allowed DM mass range.

Recall that we have not fixed the Higgsplosion scale $\EH$. Instead, in the regime considered, it is uniquely specified by the values of $\lhx$ and $\mx$ as argued in \refsec\ref{sec:model}. It is worth noting that the parameter space would be larger if we relaxed the assumption that $\lx\ll\lhx$, 
or allowed for a fine-tuning between the bare mass term and the $\lhx$ loop contribution.

Finally, we note that $\langle\sigma v\rangle$ is the cross-section at the time of freeze out, $T_f = \mx/x_f \sim 0.1 \mx$, when the $X$ species was relativistic. Therefore, the Sommerfeld enhancement \cite{PhysRevD.79.015014,Feng:2010zp} is negligible.

\medskip
\section{Constraints on the model}
\label{DMD}
\subsection{Direct detection}
\label{sec:direct}

We are now in a position to calculate the cross-sections expected in direct detection experiments, using the values of $\mx$ and $\lhx$ that are allowed by the relic density. The DM-nucleon cross-section is given by \cite{Cline:2013gha} and adapted here for the large mass limit, $\mx > m_N$):
\begin{equation}
\sigma_{SI}= \frac{\lhx^2 f_N^2}{4\pi} \frac{m_N^2 \mx^2}{(m_N + \mx)^2} \frac{m_N^2}{m_h^4 \mx^2} \sim  \frac{\lhx^2 f_N^2}{4\pi} \frac{m_N^4 }{m_h^4 \mx^2}, 
\label{eq:sigSI}
\end{equation}
where $m_N$ is the mass of a nucleon and $f_N \sim 0.3$ is an effective Higgs-nucleon-nucleon coupling.

\begin{figure}[t]
\begin{center}
\includegraphics[width=0.75\textwidth]{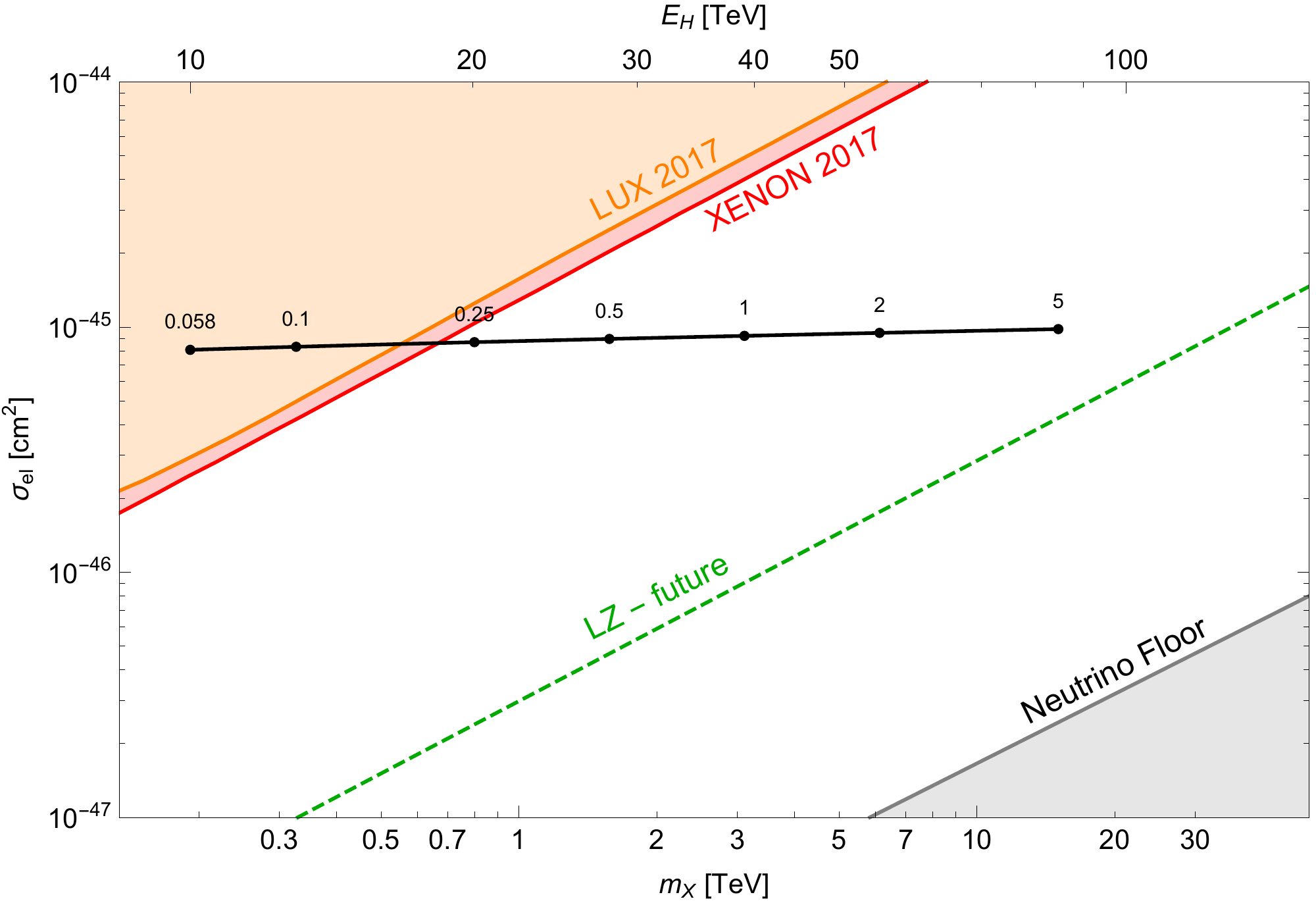}
\end{center}
\caption{Constraints on possible real singlet DM masses from present and projected future direct detection experiments. Demanding Planck relic density confines the possible values to the black line, with values of portal couplings $\lhx$ indicated for a selection of points on said line. The line terminates at the onset of non-perturbative behaviour  at $\lhx=5$. The Higgsplosion scale $E_H$ required to attain a given mass is shown on the top axis. We see that masses below $\sim 0.7$ TeV are excluded by current experiments. The future experiment, LZ, is predicted to probe the remaining perturbative mass range \cite{Mount:2017qzi}.}
\label{fig:DD}
\end{figure}

Specifying a DM mass uniquely determines $\lhx$ and $\EH$ for a Planck relic density, and so we can see how the elastic cross-section, $\sigma_{SI}$, varies with DM mass, $\mx$. This dependence is plotted in \reffig\ref{fig:DD}, with the Higgsplosion scale shown on the top axis, and the associated quartic coupling, $\lhx$, indicated for a selection of points on the line. Present and projected constraints from LUX \cite{Akerib:2016vxi}, Xenon-1T \cite{Aprile:2017iyp} and LZ \cite{Mount:2017qzi} are shown for reference. We see that current constraints exclude DM masses below $\sim 0.7$ TeV, with future searches such as LZ probing the remaining perturbative parameter space. Note that both $\sigma_{SI}$ and $\langle\sigma v\rangle$ are proportional to $\lhx^2/\mx^2$. Hence, maintaining the correct relic density keeps $\sigma_{SI}$ constant, up to the logarithmic corrections from $x_f$.

We now shift our focus to indirect detection.

\subsection{Indirect detection: observation of galactic centre by HESS}
\label{sec:indirect}

In this section we compare the constraints on annihilation cross-section derived from ten years of observations of the inner $300$ pc of the galactic centre region by the HESS experiment \cite{Lefranc:2016srp} with cross-section values predicted for our DM scalar. This involves looking at the expected photon spectrum for the processes $XX\to \{hh,W^+W^-,\tau\tau\}\to \gamma\gamma$, which we obtain using Cirelli's data and associated Mathematica package \cite{Cirelli:2010xx}.

The HESS data is presented in \cite{Lefranc:2016srp} as constraints on $\langle\sigma v\rangle$ for $XX\to \bar{J}J$, where $J\in\{W,\tau\}$. Unfortunately, our scalars annihilate into $hh$, $W_LW_L$ and $Z_LZ_L$. However, the supplied information is sufficient for a conservative bound. The rationale used in our recast is that the number of photons predicted in any bin should not exceed the number of observed photons in this bin. The conservative upper bound can be related to the constraints on the $WW$ and $\tau\tau$ channels. For a more in-depth explanation of the procedure, see Appendix~\ref{app:HESS}. We find that,
\begin{equation}
\langle \sigma v \rangle (XX\to \mathrm{SM})  \leq \min_{E \in \left[10\mathrm{GeV},\:30\mathrm{ TeV}\right]} \left[\frac{4\max\left(\sigma_{WW}\frac{dn_{WW}}{dE} ,\sigma_{\tau\tau} \frac{dn_{\tau\tau}}{dE}\right)}{\left(\frac{dn_{hh}}{dE} + 2\frac{dn_{W_LW_L}}{dE} + \frac{dn_{Z_LZ_L}}{dE} \right)}\right],
\end{equation}
where $\sigma_{WW}$ and $\sigma_{\tau\tau}$ are the HESS constraints on DM annihilating dominantly into $WW$ and $\tau\tau$ channels, and $dn_{ii}/dE$ are the spectra of photons from annihilations $XX \to ii$. 
Since the annihilations are non-relativistic, the total cross-section computed at leading order should be corrected to account for a potentially large effect due to
multiple $t$-channel exchanges of the light Higgs bosons between the non-relativistic $X$ scalars, $m_X \gg m_h$.
This gives rise to the multiplicative Sommerfeld enhancement factor $\mathcal{S}$ for the total cross-section. However, the effective coupling for $XX\to XX$ scattering through the t-channel Higgs is not $\alpha \equiv \lx^2/4\pi$, but instead a much smaller,
\begin{equation}
\alpha \equiv \left(\frac{v_{EW}}{\mx}\right)^2\frac{\lx^2}{4\pi}.
\end{equation}
This can be obtained directly by extracting the pre-factor of the Yukawa potential obtained by matching the amplitude for the process $XX \to XX$. As a result, when we use the analytic approximation (obtained from the Hulth\'en potential) for $\mathcal{S}$ given in \cite{Feng:2010zp},
\begin{equation}
\mathcal{S}=\frac{\pi}{\epsilon_v}\frac{\sinh\(\frac{2\pi\epsilon_v}{\pi^2\epsilon_h/6}\)}{\cosh\(\frac{2\pi\epsilon_v}{\pi^2\epsilon_h/6}\)-\cos\(2\pi\sqrt{\frac{1}{\pi^2\epsilon_h/6}-\frac{\epsilon_v^2}{(\pi^2\epsilon_h/6)^2}}\)},
\label{eq:sommer}
\end{equation}
we need to choose $\epsilon_v\equiv 4\pi v/\lambda_{HX}^2 (\mx/v_{EW})^2$ and $\epsilon_h \equiv 4\pi \mx m_h/(v_{EW}^2\lambda_{HX}^2)$ with velocity $v =10^{-4}c\sim 30 \mathrm{km/s}$. For this reason, we find a very minimal Sommerfeld enhancement for the range of DM masses considered, despite the non-relativistic velocity.
Finally, we also find that the enhancement factor in the relevant mass range is largely unchanged by up to order-of-magnitude changes in the velocity $v$.

\begin{figure}[t]
\begin{center}
\includegraphics[width=0.85\textwidth]{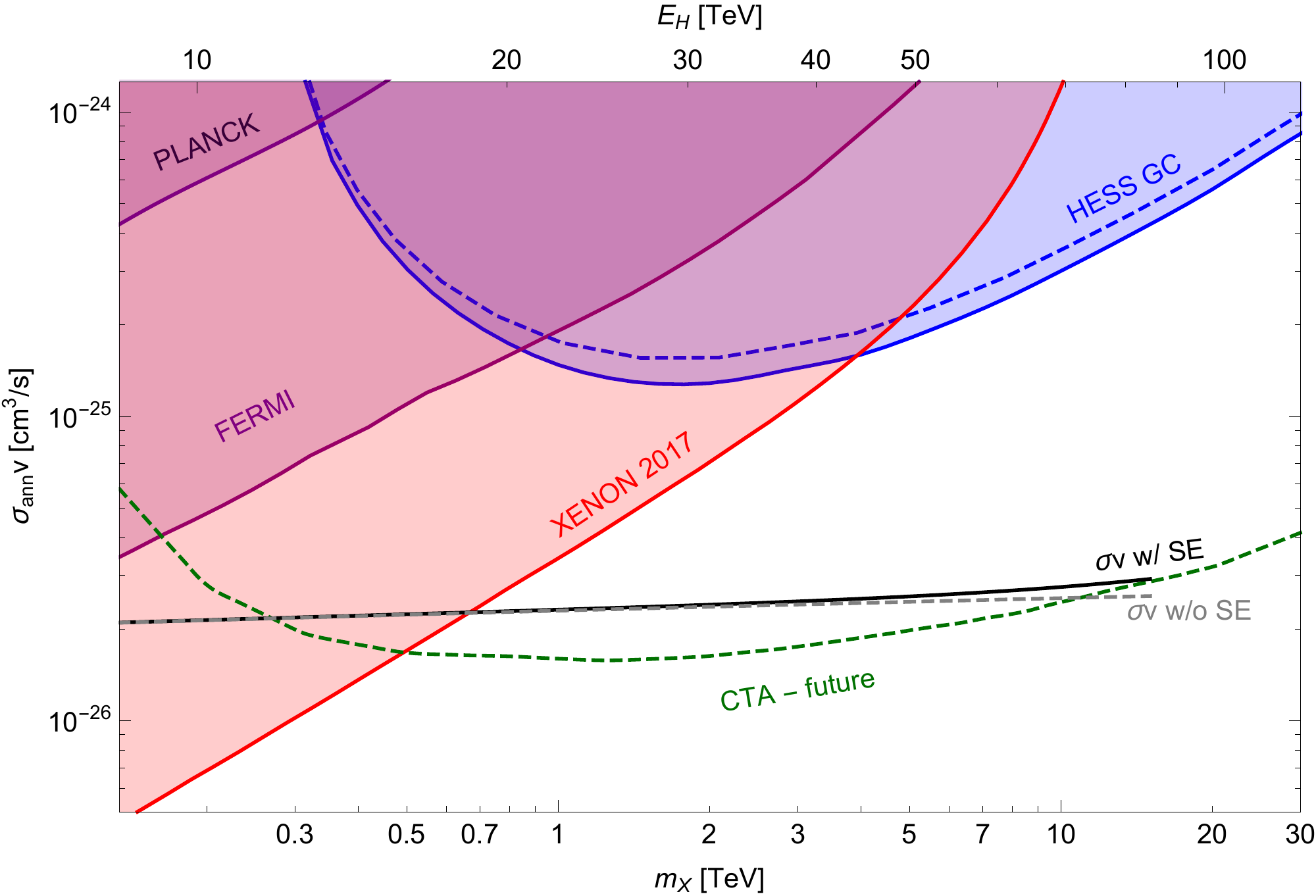}
\end{center}
\caption{Predicted theoretical annihilation cross-sections $\langle \sigma v \rangle$ for $XX\to\gamma\gamma$ in the galactic centre (GC) compared to indirect detection constraints from PLANCK, FERMI and HESS data, as well as the direct detection constraint from XENON 2017, reinterpreted in terms of $\langle \sigma v \rangle$. Shaded regions are excluded. The solid blue line indicates the recast HESS constraint which is conservative and the dashed blue line shows the constraint on pure WW decay channel. The real constraint will lie somewhere between the two, but fortunately the difference is minimal. The predicted cross-section is shown with and without Sommerfeld enhancement, up to the DM mass corresponding to a non-perturbative portal coupling, $\lhx=5$ . DM particles in the GC are expected to be non-relativistic, so enhancement considerations are necessary. We see that for the DM masses allowed by a perturbative coupling, the enhancement is minimal. The projected future constraint from CTA is shown to probe the remaining perturbative parameter space. }
\label{fig:indirect}
\end{figure}

Both the original and recast HESS results are shown in \reffig\ref{fig:indirect}, which will be discussed in \refsec\ref{sec:summ}, once all potentially competitive detection methods have been addressed.

\subsection{Other indirect searches }
\label{sec:indirectother}
Ultimately, we will see that HESS provides the strongest current indirect detection constraints on our model. However, it is worth discussing other current searches, as well as projected future constraints:

\begin{itemize}
\item The Fermi Large Area Telescope records gamma-ray emission from dwarf spheroidal galaxies, which can be used to derive constraints on the annihilation cross-section, $\langle\sigma v\rangle$, \cite{Ahnen:2016qkx,Cline:2013gha,2011PhRvL.107x1303G,Ackermann:2011wa,Tsai:2012cs}. 
\item The Planck satellite measures CMB anisotropy, which once again can be used to calculate constraints on the annihilation cross-section \cite{Ade:2015lrj,Kawasaki:2015peu}.
\item The Cherenkov Telescope Array (CTA) is a future ground-based gamma-ray observatory, which among other things, will observe the galactic centre. We use the projected contraints computed by \cite{Lefranc:2015pza}.
\end{itemize}
The constraints from the above experiments on the annihilation cross-section are shown along with the HESS and XENON constraints in \reffig\ref{fig:indirect}, which will be discussed in \refsec\ref{sec:summ}.

\subsection{Production at the LHC}
\label{sec:colliders}

In general, this class of models is very hard to discover at colliders: full production cross-section at the LHC is suppressed for three reasons: 1) Since the final state $XX$ is invisible we need to recoil against additional radiation, 2) the $XX$ state is only accessible through weak couplings in the VBF scenario or through an off-shell Higgs, 3) the mass of $X$ is large enough to pose kinematic suppression. As a result, according to \cite{Craig:2014lda}, at a $100$ TeV collider, even with $30 \rm{ab}^{-1}$, we become $1\sigma$ sensitive to $\lhx \sim 1$ for $\mx \lesssim 200$ GeV. This is well outside of the region we consider in this work.

\subsection{Summary of constraints}
\label{sec:summ}

The constraints imposed by the various direct and indirect searches described in the previous sections on annihilation cross-section are compared to the theoretical prediction (with and without Sommerfeld Enhancement) in \reffig\ref{fig:indirect}. 
In order to include the direct detection constraints from XENON 2017 on this plot, we have re-interpreted
the XENON 2017 exclusion contour for $\sigma_{SI}$ in terms of the corresponding annihilation cross-sections, $\langle \sigma v \rangle$, using \refeq\eqref{eq:sigSI}.

We see that, ultimately, HESS and XENON 2017 provide the strongest current exclusions. The Planck and Fermi results are thus omitted from later plots in the interest of clarity. The dashed blue line indicates the original $WW$ HESS data, with the solid blue line showing the conservative recast. The real constraint will lie somewhere between the two, but fortunately the difference is minimal.

\begin{figure}[t]
\begin{center}
\includegraphics[width=0.85\textwidth]{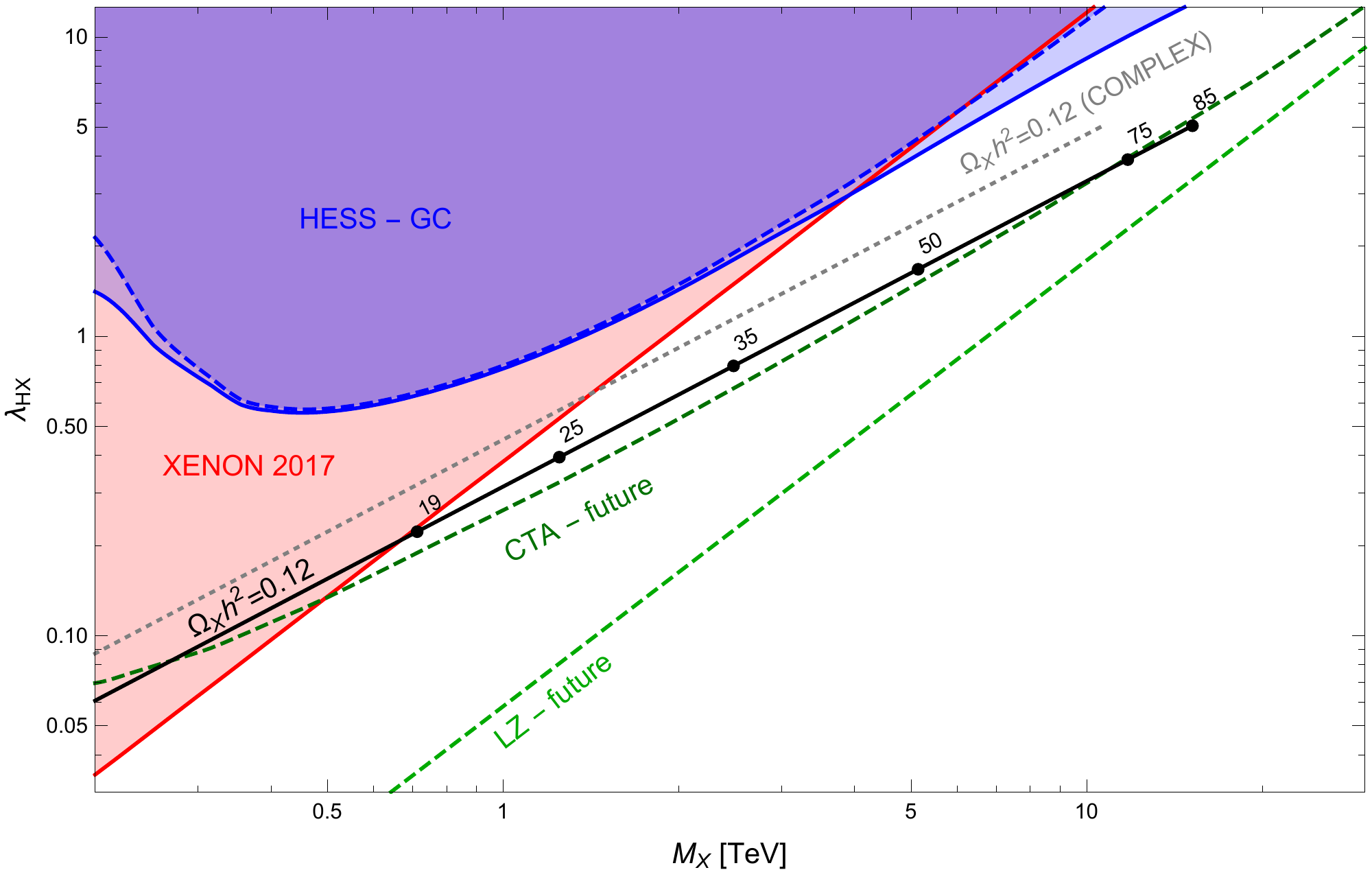}
\end{center}
\caption{Constraints from direct and indirect detection, in mass-coupling space. Demanding correct relic density confines the possible values to the black line, with numbered points indicating the required Higgsplosion scale $E_H$ in TeV. We end our line at a coupling of $\lhx=5$, which we consider non-perturbative. The grey dashed line traces the same observed value of the relic density but for the case of a complex scalar. The HESS constraints are shown with (solid blue) and without (dashed blue) Sommerfeld enhancement. We see a small effect in the regions of higher couplings. For $X$ real and perturbative coupling, we are left with a range of possible DM masses and the corresponding $E_H$ values in the range $19\, {\rm TeV} \lesssim E_H\lesssim 85\, {\rm TeV}$, that are experimentally viable. As in earlier plots, the projected future constraints from LZ and CTA probe the entire remainig parameter space.}
\label{fig:master}
\end{figure}

The region of interest is shown in $(\mx,\lhx)$ space in \reffig\ref{fig:master}. The numbered points on the black critical density line indicate the Higgsplosion scale required for that mass and coupling in TeV. We thus find ourselves with the following ranges (which are of course coupled):
\begin{equation}
\begin{aligned}
0.7~\mathrm{ TeV} & \,\lesssim \mx &\lesssim \,& 15~\mathrm{ TeV}  \\
19~\mathrm{ TeV}&  \,\lesssim  \EH &\lesssim \, & 85~\mathrm{ TeV} \\
0.2&\,\lesssim  \lhx &\lesssim\,  & 5
\end{aligned}
\end{equation}
The upper bound to this range, at least currently, comes from the requirement of a perturbative coupling. The point at which the coupling becomes non-perturbative is ill-defined and somewhat subjective: as such, so is the upper bound. We choose a maximum coupling of $\lhx =\sqrt{8\pi}\simeq5$. The lower bound is given by the XENON 2017 direct detection constraints and is far-better defined.

Conveniently, this range is accessible by detectors currently in development, such as CTA and LZ \cite{Lefranc:2015pza,Mount:2017qzi}. The model considered in this work is of course very minimal, but it is interesting that it is testable in the next decade. The dashed grey line indicates the masses and couplings required for correct relic density in the case that $X$ is a complex scalar. We see that this scenario gives a similar but slightly stricter range.
 
\begin{figure}[t]
\begin{center}
\includegraphics[width=0.85\textwidth]{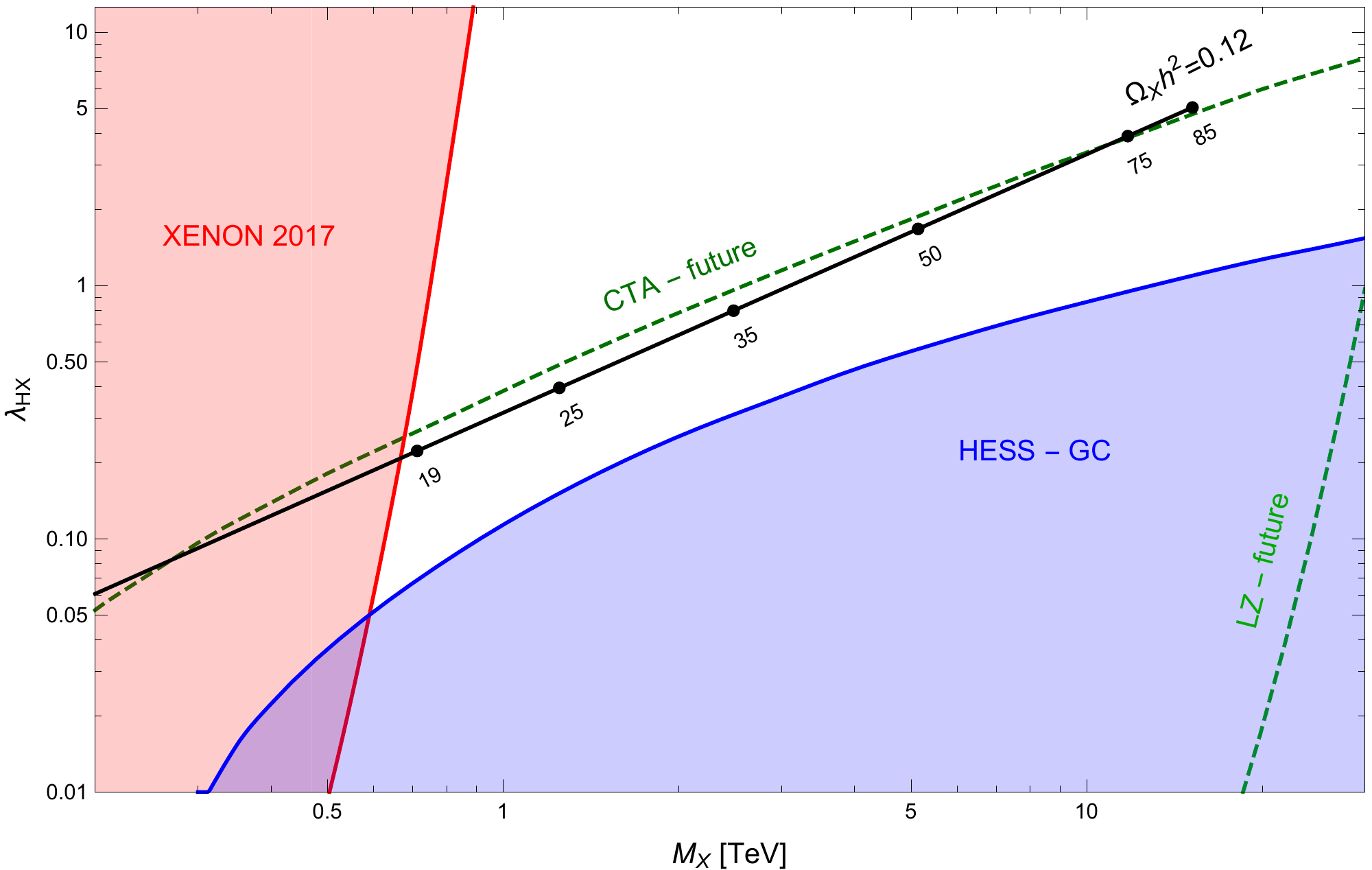}
\end{center}
\caption{Constraints from direct and indirect detection, in mass-coupling space. However, constraints are now derived without assuming correct relic density: i.e. given a relic density dictated by the $(\mx,\lhx)$ values as free independent parameters. As in \reffig\ref{fig:master}, demanding correct relic density confines the possible values to the black line, with numbered points indicating the required Higgsplosion scale in TeV. We once again end our line at a coupling of $\lhx=5$, which is considered non-perturbative. Note that for this plot, the projected exclusion zone for CTA is below its line, whereas the projected exclusion zone for LZ is above.}
\label{fig:notrelic}
\end{figure}

\medskip
The bounds in \reffig\ref{fig:master} were calculated under the assumption that the scalar DM was the sole contributor to the correct relic density, $\Omega h^2=0.12$. However, in the interest of completeness, \reffig\ref{fig:notrelic} shows the same bounds but derived using a relic density dictated by $\mx$ and $\lhx$ as independent parameters. Essentially, the direct detection limits are rescaled by $\alpha$ and indirect by $\alpha^2$, where $\alpha(\mx,\lhx) = \Omega h^2(\mx,\lhx)/0.12$. Unsurprisingly, the XENON and HESS curves still cross the black line of correct relic values at the same points, as this is where $\alpha=1$. Once again, LZ and CTA will be able to comfortably probe the entire parameter space predicted in our model.

\medskip
\section{Other considerations}
\label{sec:other}
\subsection{DM self-interaction}
\label{sec:dmsi}
In this section we assess the extent of DM self-interaction in our model. Recent work has suggested that strong DM self-interactions can change inner-halo structure and give better agreement with short-scale observations than regular cold dark matter (CDM) models. To solve these small-scale discrepancies, such as the ``core-cusp" problem, we require a self interaction cross-section per mass of $\sigma(XX\to XX)/\mx \sim 1$cm$^2$g$^{-1}$ \cite{Huo:2017vef,Tulin:2017ara}.

Self scattering is determined by two processes: $XX \to XX$, with amplitude proportional to $\lx$, and $XX\to h^* \to XX$, with amplitude proportional to $\lhx^2 v^2/m_h^2$. Since we require $\lx\ll\lhx$ in our model, the Higgs exchange process dominates the amplitude. Summing the contributions from the s, t and u channels and calculating the cross-section yields a non-trivial expression. However, for $m_h\ll \mx$ and under the assumption that that the initial and final $X$ particles are non-relativistic, $v\ll 1$, we find the leading order term,
\begin{equation}
\sigma(XX\to XX) = \frac{v_{EW}^4\lhx^4}{16m_h^4\mx^2}+\mathcal{O}(v^2,m_h^{-2}),
\end{equation}
where $v$ is the velocity, whereas $v_{EW}$ is the Higgs VEV as before. This corresponds to a cross-section per mass of order
\begin{equation}
\begin{split}
\frac{\sigma(XX\to XX)}{\mx}&\approx \lhx^4\(\frac{\mx}{1\mathrm{ TeV}}\)^{-3}(2.1\times 10^{-15}\mathrm{cm}^2\mathrm{g}^{-1})\\
&\approx 10^{-18}\mathrm{cm}^2\mathrm{g}^{-1}\; \(\frac{x_f}{20}\)^{2}
\(\frac{\Omega_\mathrm{X}h^2}{0.12}\)^{2}
\(\frac{\mx}{1\mathrm{ TeV}}\) .
\end{split}
\end{equation}
As a result, even with large Sommerfeld enhancement, we do not expect to see any significant self-interaction for this dark matter model.

\subsection{Vacuum stability and the RG flow of quartic couplings}
\label{sec:vac}

In this section we will show that the vacuum of our model is stable at the tree level and that this property is unchanged by renormalisation group (RG) flow.
The scalar potential in our theory is given by,
\begin{equation}
V(A,B) = -\mu_0^2A+\frac{1}{2}m_{X,0}^2B+\lh A^2+\frac{\lx}{4!}B^2 + \frac{\lhx}{2}AB,
\end{equation}
where $A = H^\dagger H>0$ and $B =X^2>0$. Vacuum stability is achieved if either of the following criteria is satisfied:
\begin{equation}
\begin{aligned}
\mathrm{if}&\quad \lhx<0:\quad &\quad \lh,\lx>0&\quad \mathrm{and} \quad \lh\lx> \frac{3}{2}\lambda^2_{HX},\\
\mathrm{if}&\quad \lhx>0:\quad &\quad \lh,\lx>0 .&
\end{aligned}
\end{equation}

The case of negative $\lhx$ is in tension with our earlier assumption of $\lhx\gg\lx$, because negative $\lhx$ implies $\lh\lx> \frac{3}{2}\lhx^2$. However, if all three couplings are positive, the vacuum is stable even for $\lhx\gg\lx$. In both scenarios, the vacuum is most easily destabilised if $\lx$ runs to a negative value.

Since we require $\lx \ll \lhx \sim 1$, a small change to $\lx$ due to RG flow can change the sign of $\lx$ with catastrophic consequences. We require small $\lx$ at RG scale $\mu = \EH$, in order to avoid additional contribution to $\mx$. Above $\EH$ the RG flow is frozen due to the 
Higgsplosion of the self-energy leading to the Higgspersion of full propagaotrs \cite{Khoze:2017tjt,Khoze:2017lft,Khoze:2017uga}. Hence, we need only worry about the small region between $\mx$ and $\EH$. In this region, the approximate change to $\lx$, $\delta \lx$, is simply,
\begin{equation}
\delta \lx \sim \beta_{\lx} \log \left(\frac{\mx}{\EH}\right) \sim - \frac{3}{16\pi^2} \lhx^2 \log\left(\frac{4\pi}{\sqrt{\lhx}}\right),
\end{equation}
where $\beta_{\lx}$ is the beta function for the quartic $X$ coupling (for full expressions see Appendix~\ref{app:RG}).
As long as $\lx(\EH) > |\delta \lx|$, then the vacuum is stable. As a result we require:
\begin{equation}
\lx \gtrsim \frac{3\lhx^2}{16\pi^2}  \log\left(\frac{4\pi}{\sqrt{\lhx}}\right).
\end{equation}

For the two extremal allowed values of  $\lhx=0.2$ and $\lhx = 5$, this corresponds to $\lx \gtrsim 2.5\times 10^{-3}$ and $\lx \gtrsim 0.82$ respectively. As a result, it is always possible to achieve at least $\lx / \lhx \lesssim 0.17$, and so our assumption, $\lx\ll\lhx$, is safe (particularly for lower DM masses). We perform a more accurate analysis with the full coupled evolution of all three couplings $\{\lx,\lhx,\lh\}$ in Appendix~\ref{app:RG}. The conclusions remain essentially the same: we can achieve vacuum stability at all scales, while maintaining the core assumption of this work, $\lx\ll\lhx$.

\section{Discussion and conclusions}
\label{sec:concl}
\subsection{Relaxing assumptions and extending the dark sector}
In this paper we have made the following assumptions:  
\begin{enumerate}
\item There is no fine-tuning in the mass of $X$. Furthermore, the radiative correction to the mass of $X$ is determined by corrections from the Higgs particle.
\item $X$ is the sole component of a dark sector.
\end{enumerate}

If we were to relax these assumptions, the constraints on this model would be weakened. Consider the following examples:

\begin{enumerate}
\item Breaking the first assumption has simple consequences. Requiring the correct relic density still fixes a relationship between $\lhx$ and $\mx$. However, allowing for additional contributions, $\mu_\mathrm{X}^2$, to the DM mass relaxes the relationship between $\lhx$, $\mx$ and $\EH$,
\begin{equation}
\mx^2 = \mu_\mathrm{X}^2 + \lhx\frac{\EH^2}{16\pi^2}.
\end{equation}
As a  result, for negative $\mu_\mathrm{X}^2$ we can obtain the same combination of $\lhx$ and $\mx$ for a higher value of $\EH$. From the point of view of physics below the Higgsplosion scale, nothing is influenced by this fine-tunning. 
\item Breaking the second assumption can be done in many ways and the phenomenological implications are just as varied. For example, we can introduce another species that $X$ can freeze-out into. Consequently, the annihilation cross-section for $XX$ may become independent of $\lhx$ and completely ruin the predictive power of this model: we can set $\lhx$ arbitrarily small and make $X$ invisible while maintaining the correct relic density. For another example of breaking the second assumption, we refer the reader to \cite{Casas:2017jjg}.

\end{enumerate}

Though all these changes to our model are reasonable, they do lead to less predictive and more complicated scenarios. We therefore refrain from exploring such extensions.

\subsection{Conclusions}

In this paper we have shown that the Higgsplosion mechanism results in a definite prediction for mass of a real scalar dark matter candidate, as well as a definite prediction of its coupling to the Standard Model fields. In particular, the lowest value for the Higgsplosion scale theoretically preferred, i.e. $\EH \sim 25$ TeV,  implies a dark matter mass of $\mx \sim 1.25$ TeV and a Higgs portal coupling $\lhx \sim 0.4$, which remains safe from all current constraints.

In order to check the viability of this scenario, we have updated the direct detection constraints on scalar dark matter models for the newest dataset from the XENON experiment, and have recast recent HESS indirect detection results. 

This particular model can be probed by future indirect detection experiments such as CTA, and as well as by the direct detection experiment LZ, both of which should start collecting data within the next decade. As a result, we have presented a definite prediction for a model of dark matter that can be, in its present form, discovered in the foreseeable future. It is possible to relax some of the (fairly strict) assumptions to relax bounds on this model, at the cost of loss of predictivity. We invite anyone interested to do so.

\bigskip

\section*{Acknowledgements}
We would like to thank Mikael Chala, Richard Ruiz and Matthew Kirk for helpful discussions.

\newpage
\appendix
\appendixpage
\addappheadtotoc

\section{Annihilation cross-sections}
\label{app:xsec}

Below are the cross-sections for annihilation modes as calculated in \cite{PhysRevD.50.3637}:
\begin{align}
\langle \sigma_{hh}v\rangle&=\frac{\lhx^2}{64\pi \mx^2}\(1-\frac{m_h^2}{\mx^2}\)^{1/2}& &\xrightarrow{\mx^2\gg m_h^2}& &\frac{\lhx^2}{64\pi \mx^2},\notag\\
\langle \sigma_{WW}v\rangle&=\frac{\lhx^2m_W^4[2+(1-2\mx^2/m_W^2)^2]}{8\pi \mx^2[(4\mx^2-m_h^2)^2+m_h^2\Gamma_h^2]}\(1-\frac{m_W^2}{\mx^2}\)^{1/2} & &\xrightarrow{\mx^2\gg m_h^2,m_W^2}& &\frac{\lhx^2}{32\pi \mx^2},\notag\\
\langle \sigma_{ZZ}v\rangle&=\frac{\lhx^2m_Z^4[2+(1-2\mx^2/m_Z^2)^2]}{16\pi \mx^2[(4\mx^2-m_h^2)^2+m_h^2\Gamma_h^2]}\(1-\frac{m_Z^2}{\mx^2}\)^{1/2} & &\xrightarrow{\mx^2\gg m_h^2,m_Z^2}& &\frac{\lhx^2}{64\pi \mx^2},\notag\\
\langle \sigma_{ff}v\rangle&=\frac{\lhx^2m_f^2}{4\pi[(4\mx^2-m_h^2)^2+m_h^2\Gamma_h^2]}\(1-\frac{m_f^2}{\mx^2}\)^{3/2}& &\xrightarrow{ \mx^2\gg m_h^2,m_f^2}& &\frac{\lhx^2}{64\pi \mx^2}\frac{m_f^2}{\mx^2}.
\end{align}
We see that in the limit $\mx^2 \gg m_h^2, m_W^2, m_Z^2, m_f^2$, the total annihilation cross-section is:
\begin{equation}
\langle \sigma v \rangle = \frac{\lhx^2}{16\pi \mx^2} + \mathcal{O}\(\frac{m_h^2}{\mx^2},\frac{m_W^2}{\mx^2}, \frac{m_Z^2}{\mx^2},\frac{m_f^2}{\mx^2}\)
\end{equation}

\section{HESS data recast}
\label{app:HESS}

The HESS instrument has measured the high energy photon spectrum over the last 10 years (254 hours of observation) \cite{Lefranc:2016srp}. Unfortunately, there are only two dark matter annihilation channels whose cross-sections have been officially constrained by the HESS data: $XX \to W^+W^-$ and $XX \to \tau^+ \tau^-$. We will reinterpret these results in order to derive a conservative constraint on a combination $XX \to H^\dagger H$, which is a weighted combination of three channels $XX \to hh$, $XX \to W_L^+W_L^-$ and $XX\to Z_LZ_L$.

First we define several variables: let $\mathcal{L}$ be the line of sight integral,
\begin{equation}
\mathcal{L} = \int n^2 dl,
\end{equation}
where $n$ is the dark matter density, such that $\mathcal{L} \langle \sigma v \rangle$ gives the rate of dark matter particle annihilation along the line of sight. Note that $\mathcal{L}$ only depends on the dark matter density along the line of sight and so it is independent of the annihilation channel we are constraining.

Furthermore, we define the spectral density of photons from a single annihilation of two dark matter particles into a final state J as $dn_{J}/{dE}$. For example, the spectrum of photons from the process $XX \to W_LW_L$ is denoted by $dn_{W_LW_L}/dE$. Note that the integral,
\begin{equation}
N_\gamma = \int_{E_0}^{\infty} \frac{dn_J}{dE} dE = \langle N_\gamma(E>E_0) \rangle,
\end{equation}
is the average number of photons with energy larger than $E_0$ from a single annihilation $XX\to J$, and therefore is not bounded by 1. We used Cirelli's data and associated Mathematica package \cite{Cirelli:2010xx}, which greatly simplified our work. 

Now that we are ready, consider an energy bin $[E_1,E_2]$. The conservative constraints on the number of photons in this bin $N_c$, as inferred from the cross-section constraints ($\sigma_{WW}$ and $\sigma_{\tau\tau}$) on these channels, are:
\begin{align}
N_{c,WW}(E_1,E_2) &= \mathcal{L}\sigma_{WW}\int_{E_1}^{E_2} \frac{dn_{WW}}{dE} dE,\\
N_{c,\tau\tau}(E_1,E_2)&= \mathcal{L}\sigma_{\tau\tau}\int_{E_1}^{E_2} \frac{dn_{\tau\tau}}{dE} dE.
\end{align}
Therefore, the conservative upper bound on the number of observed photons in this bin is:
\begin{align}
N_{c}(E_1,E_2) = \max \left(N_{c,WW}(E_1,E_2) ,N_{c,\tau\tau}(E_1,E_2)  \right).
\end{align}
On the other hand, given the annihilation cross-section, $\sigma_{ann} = \sigma(XX \to H^\dagger H)$, we can compute the predicted number of photons in this bin:
\begin{equation}
N_p(E_1,E_2) = \mathcal{L}\int_{E_1}^{E_2} \frac{\sigma_{ann}}{4} \left(\frac{dn_{hh}}{dE} + 2\frac{dn_{W_LW_L}}{dE} + \frac{dn_{Z_LZ_L}}{dE} \right) dE.
\end{equation}
Given that we want to constrain $\sigma_{ann}$, we require that,
\begin{equation}
N_p(E_1,E_2) \leq N_c(E_1,E_2), \label{eq:ineq}
\end{equation}
for all choices of $(E_1,E_2)$ within the observational range of HESS: 10 GeV -- 30 TeV. This means that the inequality in \refeq(\ref{eq:ineq}) has to be satisfied at the integrand level,
\begin{equation}
\frac{\sigma_{ann}}{4} \left(\frac{dn_{hh}}{dE} + 2\frac{dn_{W_LW_L}}{dE} + \frac{dn_{Z_LZ_L}}{dE} \right) \leq \max\left(\sigma_{WW}\frac{dn_{WW}}{dE} ,\sigma_{\tau\tau} \frac{dn_{\tau\tau}}{dE}\right),\;\forall E \in [10\mathrm{GeV},30\mathrm{ TeV}].
\end{equation}
Finally, this implies that the bound on $\sigma_{ann}$ is:
\begin{equation}
\sigma_{ann}  \leq \min_{E \in \left[10\mathrm{GeV},\:30\mathrm{ TeV}\right]} \left[\frac{4\max\left(\sigma_{WW}\tfrac{dn_{WW}}{dE} ,\sigma_{\tau\tau} \tfrac{dn_{\tau\tau}}{dE}\right)}{\left(\tfrac{dn_{hh}}{dE} + 2\tfrac{dn_{W_LW_L}}{dE} + \tfrac{dn_{Z_LZ_L}}{dE} \right)}\right].
\end{equation}

\section{More on the RG flow of quartic couplings}
\label{app:RG}

Our $\beta$-functions, $\beta_{\lambda_i}$, have the usual definitions,
\begin{equation}
\beta_{\lambda_i}(\lh,\lx,\lhx)=\frac{d\lambda_i}{d\log\mu},
\end{equation}
where $\lambda_i\in \{\lh,\lx,\lhx\}$ and $\mu$ is some RG scale. Unsurprisingly, we see similarities with $\phi^4$ theory, where $\beta(\lambda) = 3\lambda^2/16\pi^2$. Here, we specifically find,
\begin{equation}
\begin{aligned}
\beta_{\lh}(\lh,\lx,\lhx) &\approx \frac{3}{96\pi^2}(36\lh^2+\lhx^2)+\mathcal{O}(\lambda_i^3)\\
\beta_{\lx}(\lh,\lx,\lhx) &\approx \frac{3}{16\pi^2}(\lx^2+\lhx^2)+\mathcal{O}(\lambda_i^3)\\
\beta_{\lhx}(\lh,\lx,\lhx) &\approx \frac{1}{16\pi^2}(6\lh\lhx+\lx\lhx+4\lhx^2)+\mathcal{O}(\lambda_i^3)
\end{aligned}.
\end{equation}
Note that the slightly differing pre-factors in the expressions above are merely an artefact of the way $\lh$ is normalised in the conventional SM, versus the more canonical way in which $\lx$ and $\lhx$ have been normalised.
 
As explained in \refsec\ref{sec:vac}, to ensure vacuum stability we require that $\lx$ does not become negative when RG running down from the Higgsplosion scale, $\EH$, to the DM mass, $\mx$. Above $\mu=\EH$, the RG flow is frozen by Higgsplosion. This is also the scale at which we require that $\lx \ll \lhx$ in order to prevent a significant contribution to $\mx$.  

Suppose we set our new non-SM couplings at the Higgsplosion scale, $\mu=\EH=\EH(\mx)$, as follows,
\begin{equation}
\lh(v_{EW})=1/8\qquad\lx(\EH)=\hat\lambda_{\mathrm{X}}\qquad\lhx(\EH)=16\pi^2\(\dfrac{\mx}{\EH(\mx)}\)^2,
\end{equation}
where $\lhx(\EH)$ is fully determined by our choice of DM mass, the relation in Eq. \ref{eq:approx} and the requirement for correct relic abundance. For now, $\hat\lambda_{\mathrm{X}}$ is left as a free parameter. 

Any running derived from the $\beta$ functions above is valid down to $\mu=\mx$. Hence, we now ask, how valid is our approximation, $\lx\ll\lhx$, if we demand vacuum stability? In other words, what is the lowest $\hat\lambda_{\mathrm{X}}$ one can start with and not run into negative values for scales $\mx<\mu<\EH$? We focus on the two extremal non-excluded cases found in \refsec\ref{sec:summ}: $(\mx/\mathrm{TeV},\lhx,\EH/\mathrm{TeV})=(0.7,0.2,19)$ and $(15,5,85)$.

For the lower DM mass bound, $\mx=0.7$ TeV, we find $\lx/\lhx\approx 1.3\times10^{-2}$, while for the upper bound, $\mx=15$ TeV, we find $\lx/\lhx\approx 0.17$. So we see that our assumption, $\lx\ll\lhx$, becomes less applicable as the portal coupling, $\lhx$, increases. However, even at the largest coupling, one still finds a fairly minimial ratio of $\approx17 \%$. Therefore, we conclude that any contributions from the quartic $X$ vertex are too small to change any phenomenological aspects of the model presented in this work: vacuum stability at all scales is certainly achievable.

\newpage
\bibliographystyle{JHEP}
\bibliography{XDM_Ref}

\end{document}